\documentclass{ifacconf}
\usepackage{natbib}        
\usepackage{color,soul}
\usepackage{epstopdf}
\usepackage{multicol}
\usepackage{amssymb, amsmath, amsfonts}
\usepackage{epstopdf}
\usepackage[pdftex]{graphicx}
\usepackage{algorithm,  multirow} 
\usepackage[T1]{fontenc}

\usepackage{tikzsymbols}
\DeclareMathAlphabet{\mathbbold}{U}{bbold}{m}{n}

\usepackage[english]{babel}
\usepackage{amsmath,amssymb,amsfonts}
\usepackage{xcolor}
\usepackage{tabularx}
\usepackage{booktabs}
\usepackage{multirow}
\usepackage{courier}
\usepackage{float}
\usepackage{makecell}
\setcellgapes{5pt}
\newcommand{\ra}[1]{\renewcommand{\arraystretch}{#1}}

\newcommand*{\argminOp}{\operatornamewithlimits{argmin}\limits}

\newcommand{\tr}{{\mathsf{T}}}

\iftrue

\else

\fi

\newcommand{\vc}[1]{{ \mathrm{#1} }}

\newcommand{\Dcal}{{\mathcal{D}}}

\newcommand{\Ncal}{{\mathcal{N}}}

\newcommand{\Xcal}{{\mathcal{X}}}

\newcommand{\Rbb}{{\mathbb{R}}}

\newcommand{\vat}{v_{\text{a}}} 
\newcommand{\Vas}{V_{\text{a}}} 

\newcommand{\iat}{i_{\text{a}}} 

\newcommand{\GP}[2]{{\mathcal{G}\!\mathcal{P}}(#1,#2)}

\newcommand{\fs}{f^{\text{(s)}}}                
\newcommand{\Fs}{F^{\text{(s)}}}                
\newcommand{\Ns}{N^{\text{(s)}}}                
\newcommand{\hs}{h^{\text{(s)}}}                
\newcommand{\Ts}{T^{\text{(s)}}_{90}}           
\newcommand{\es}{e^\text{(s)}}                  
\newcommand{\gammas}{\gamma^\text{(s)}}         
\newcommand{\fp}{f^{\text{(p)}}}                
\newcommand{\Fp}{F^{\text{(p)}}}                
\newcommand{\Np}{N^{\text{(p)}}}                
\newcommand{\hp}{h^{\text{(p)}}}                
\newcommand{\hps}{h^{\text{(p)}}_\text{s}}      
\newcommand{\Tp}{T^{\text{(p)}}_{90}}           
\newcommand{\ep}{e^\text{(p)}}                  
\newcommand{\gammap}{\gamma^\text{(p)}}         

\begin{document}
\begin{frontmatter}
\title{Cascade Control: Data-Driven Tuning Approach Based on Bayesian Optimization}
\thanks[footnoteinfo]{This paper is partially supported by the Swiss Competence Center for Energy Research SCCER FEEB\&D of the Swiss Innovation Agency Innosuisse.}
\author[First]{Mohammad Khosravi}
\author[First]{Varsha Behrunani}
\author[First]{Roy S.~Smith}
\author[First,Second]{Alisa Rupenyan}
\author[First]{John Lygeros}
\address[First]{Automatic Control Lab, ETH, Z\"{u}rich 8092, Switzerland \\ (e-mail:\ khosravm@control.ee.ethz.ch, bvarsha@student.ethz.ch, rsmith@control.ee.ethz.ch, ralisa@ethz.ch, 	 		jlygeros@ethz.ch).}
\address[Second]{Inspire AG, Z\"{u}rich 8092, Switzerland \\ (e-mail:\ rupenyan@inspire.ethz.ch).}
\begin{abstract} 
~ Cascaded controller tuning is a multi-step iterative procedure that needs to be performed routinely upon maintenance and modification of mechanical systems. An automated data-driven method for cascaded controller tuning based on Bayesian optimization is proposed. The method is tested on a linear axis drive, modeled using a combination of first principles model and system identification. A custom cost function based on performance indicators derived from system data at different candidate configurations of controller parameters is modeled by a Gaussian process.  It is further optimized by minimization of an acquisition function which serves as a sampling criterion to determine the subsequent candidate configuration for experimental trial and improvement of the cost model iteratively, until a minimum according to a termination criterion is found. This results in a data-efficient procedure that can be easily adapted to varying loads or mechanical modifications of the system. The method is further compared to several classical methods for auto-tuning, and demonstrates higher performance according to the defined data-driven performance indicators. The influence of the training data on a cost prior on the number of iterations required to reach optimum is studied, demonstrating the efficiency of the Bayesian optimization tuning method.

\end{abstract}

\begin{keyword}
PID tuning, auto-tuning, Gaussian process, Bayesian optimization
\end{keyword} 
\end{frontmatter}
\section{Introduction}
Numerous systems in manufacturing rely on linear or rotational drives, often controlled by cascaded PID loops. Tuning and re-tuning these controllers is a task that needs to be routinely performed. However, it is hard to pinpoint a standard solution for it. One of the challenges in controlling and optimizing such systems is that the controller gains change with the different loads applied to the system, depending on the operational mode, and they are also dependent on drifts in friction, or loosening of the mechanical components. Often, to avoid excessive re-tuning the controller parameters are set to conservative values compromising the performance of the system while maintaining stability for a wide range of loads or mechanical properties.

Standard methods, such as the Ziegler-Nichols rule or relay tuning with additional heuristics are routinely used for the tuning in practice. Optimization of a performance criterion such as the integral of the absolute time error (ITAE) is another possible method of auto-tuning. Such methods could be simple to apply for single loop controllers. However, the complexity and the number of parameters increases in cascade control.

We propose a data driven approach for auto-tuning of the controller parameters using Bayesian Optimization (BO). The approach has been previously explored in \citep{BayesOpt1, BayesOpt2}, and \citep{khosravi2019controller}. Here we apply it for a cascade control of a linear motion system, and compare the achieved performance with standard tuning methods. The tuning problem is formulated as optimization where the controller parameters are the variables that ensure a minimum in the cost defined through a weighted sum of performance metrics extracted from the data (encoder signals). The cost is modelled as a Gaussian process (GP), and measurements of the performance of the plant are conducted only at specific candidate configurations, which are most informative for the optimization of the cost. These candidate configurations are determined through the maximization of an acquisition function that evaluates the cost function GP model, using information about the predicted GP mean and the associated uncertainty at each candidate location. In mechatronic systems, the performance depends mostly on stability restrictions, overshoot specifications, and set point tracking specifications. In this work, we restrict the range of the optimization variables to a limited set where the system is stable and focus on overshoot and set point tracking errors.
Bayesian optimization in controller tuning where stability is guaranteed through safe exploration has been proposed in \citep{BayesOpt1}, and applied for robotic applications \citep{BayesOpt5}, and in process systems \citep{khosravi2019controller,khosravi2019machine}. The proposed Bayesian optimization tuning ensures a compromise between the need of extensive number of trials for finding the optimal gains (according to a specified performance criterion), and a single trial, as resulting from standard methods, where a sub-optimal gain with respect to the performance of the system is found, but stability is ensured for a wide range of operation. With BO tuning, a small number of experiments is sufficient to find an optimal gain.

The paper is organized as follows:  Section \ref{sec:system_model} presents the model of the linear axis actuator derived from first principles in combination with system identification techniques. Section \ref{sec:numerical} presents numerical results comparing the performance of BO tuning with standard approaches (Ziegler Nichols, ITAE tuning, and relay), as well as with a brute force result derived from evaluation of the performance metrics on a grid. A study on the required number of evaluations for estimating the prior on the cost as well as the number of BO iterations is included.
Section \ref{sec:conclusion} concludes the work.
\section{System Structure and Model} \label{sec:system_model}
\begin{figure}[ht]
	\centering
	\includegraphics[width = 0.4\textwidth]{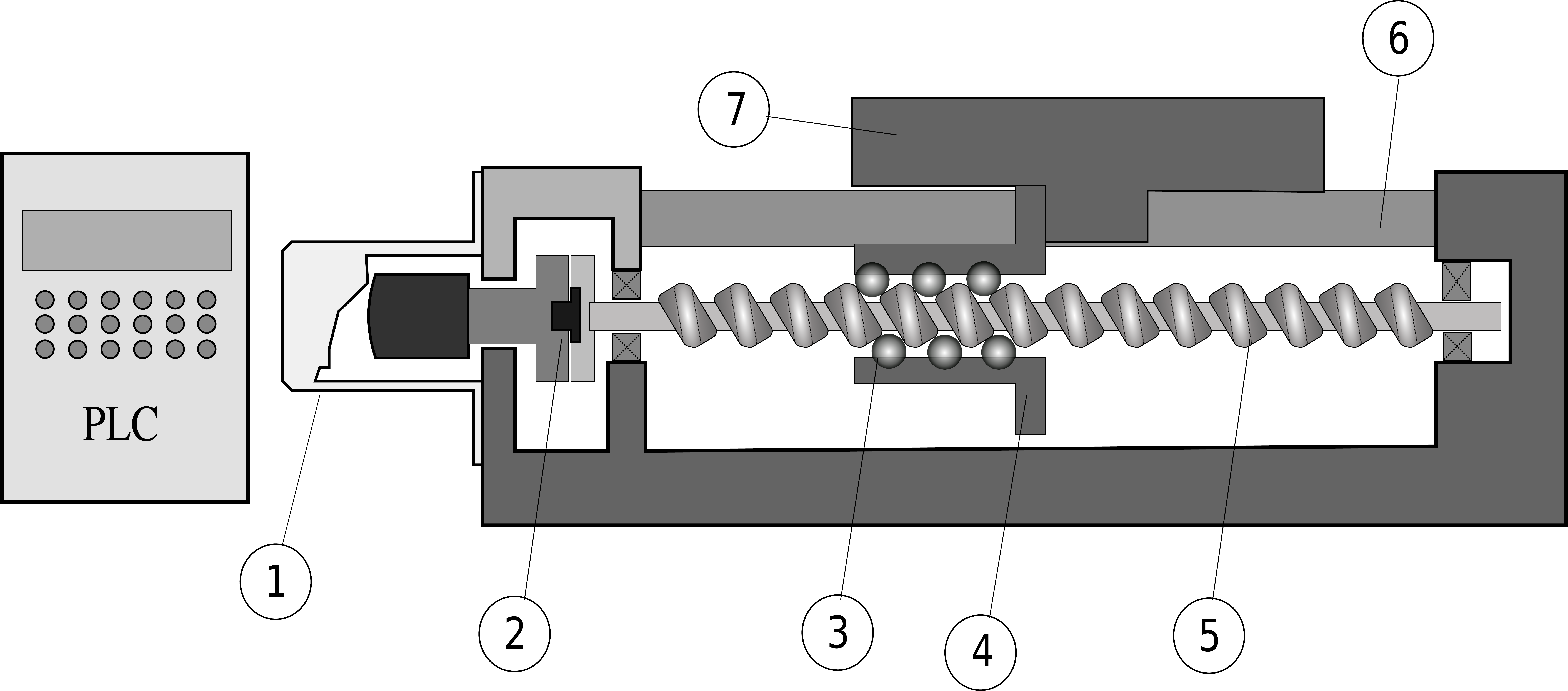}%
	\caption{The structure of ball-screw system:
	\textcircled{\small{1}} DC motor, \textcircled{\small{2}} coupling joint, \textcircled{\small{3}} ball-screw interface, 
	\textcircled{\small{4}} nut, 
	\textcircled{\small{5}} ball-screw shaft, \textcircled{\small{6}} guideway and 
	\textcircled{\small{7}} table (load), following \citep{altintas2011machine}.
	}
    \label{mfig:structure}
\end{figure}

The system consists of several units shown in Figure \ref{mfig:structure}. 
The heart of the plant is a {\em permanent magnet electric motor}. 
The motor is connected via a coupling joint to a ball-screw shaft which is fixed to a supporting frame. 
The shaft carries a nut which has a screw-nut interface converting the rotational motion of the shaft to the linear motion of the nut.
On the nut, a carriage table is fixed which can slip on the guide-ways and allows carrying loads.
The motor is actuated with a motor driver  which provides the required current and voltage to the armature of the motor. 
The motor and the ball-screw part is controlled by a PLC towards obtaining desired behavior and precision
by regulating the set voltage of the motor. The PLC takes feedback from the position and velocity of nut and also the current of the motor shaft. 
The system is equipped with encoders for measuring the position of the nut, the rotational speed of the motor and the ball-screw shaft. 
Figure \ref{mfig:structure} shows the details of the plant and connected units.\\
\subsection{Mathematical Model} \label{sec_plant_math}
To obtain a mathematical representation of the plant
\citep{ExtendedBOCascade}, we need to model the electrical and mechanical parts of the system.
We first derive the dynamics of the motor modeled as a permanent magnet DC motor using the equivalent electrical circuit and the mechanical equations of motion. 

Let $\vat$, $\iat$, $R_{\text{a}}$ and $L_{\text{a}}$ respectively denote the voltage, the current, the resistance and the inductance of the armature coils. 
From Kirchhoff's voltage law and the back electromotive force (EMF), one has 
\begin{equation}\label{eqn:v_a_dynamics}
\vat(t) = L_{\text{a}} \frac{\mathrm{d}}{\mathrm{d}t}\iat(t)+ R_{\text{a}}\iat(t) + K_{\text{b}}\omega_{\text{m}}(t) ,
\end{equation}
where $K_{\text{b}}$ is the back EMF constant and  $\omega_{\text{m}}$ is the angular velocity of the motor and shaft.
The motor develops an electromagnetic torque, denoted by $\tau_{\text{m}}$, proportional to the armature current $\tau_{\text{m}} = K_{\text{t}} \iat$.
Using Laplace transform and \eqref{eqn:v_a_dynamics}, the transfer function of motor is derived as 
\begin{equation}\label{eqn:M(s)}
M(s) 
\!:=\! 
\frac{\Omega_{\text{m}}(s)}{\Vas(s)}  
\!=\! 
K_{\text{t}}
\ \!
\big{(}K_{\text{t}}K_{\text{b}} \!+\!(L_{\text{a}} s \!+\! R_{\text{a}})\Big(\frac{T_{\text{m}}(s)}{\Omega_{\text{m}}(s)}\Big)\big{)}^{-1}
\!\!,
\end{equation}
where $\Omega_{\text{m}}$, $\Vas$ and $T_{\text{m}}$ are the Laplace transform of $\omega_{\text{m}}$, $\vat$ and $\tau_{\text{m}}$, respectively.
The main impact on the linear position is due to the {\em first axial mode} of the ball-screw system \citep{AxisModelling}, determined by the flexibility characteristics of the translating components.
The first axial dynamics of the ball-screw servo drive can be modeled using a simplified two degree of freedom mass-spring-damper system \citep{altintas2011machine}.  
Define $J_{\text{m}}$, $B_{\text{m}}$ and $\theta_{\text{m}}$ respectively as inertia of the rotor, the damping coefficient of the motor and the angular displacement of the motor.
Similarly, let $J_{\text{l}}$, $B_{l}$, $\theta_{\text{l}}$ and $\omega_{\text{l}}$ denote as 
the inertia of the load, the damping coefficient of the load, the angular displacement of the load and the angular velocity of the load, respectively.
According to the torque balance equation, we have
\begin{equation}\label{eqn:omega_ml}
\begin{array}{ll}
&\!\! J_{\text{m}} \frac{\mathrm{d}\omega_{\text{m}}}{\text{d}t} \!+\! B_{\text{m}} \omega_{\text{m}} \!+\! B_{\text{ml}} (\omega_{\text{m}} \!-\! \omega_{\text{\text{l}}}) \!+\! K_{\text{s}} (\theta_{\text{m}}\!-\! \theta_{\text{l}}) \!=\! \tau_{\text{m}},\! \\&
\!\! J_{\text{l}}\ \! \frac{\text{d}\omega_{\text{l}}}{\text{d}t} \!+\! B_{\text{l}}\ \!\omega_{\text{l}} \!-\! B_{\text{ml}} (\omega_{\text{m}}\!-\! \omega_{\text{\text{l}}}) \!-\! K_{\text{s}} (\theta_{\text{m}} \!-\! \theta_{\text{l}}) \!=\! \tau_{\text{l}},\!
\end{array} \!\!
\end{equation}
where $K_{\text{s}}$ is the axial stiffness, $\tau_{\text{l}}$ is the torque disturbance of the load and $B_{\text{ml}}$ is the damping coefficient between the coupling and the guides. 
Since $B_{\text{l}}$ has a negligible impact on resonance, one can set $B_{\text{l}}=0$ \citep{altintas2011machine}. 
Let $\Theta_{\text{m}}$, $\Theta_{\text{l}}$, $T_{\text{m}}$ and $T_{\text{l}}$ be respectively the Laplace transform of 
$\theta_{\text{m}}$, $\theta_{\text{l}}$, $\tau_{\text{m}}$ and $\tau_{\text{l}}$.
From \eqref{eqn:omega_ml}, we have
\begin{equation}\label{eqn:Transfer_theta_tau}
\begin{bmatrix}
\Theta_{\text{m}}(s)\\\Theta_{\text{l}}(s)
\end{bmatrix}
=
\mathrm{H}(s)^{-1}
\begin{bmatrix}
T_{\text{m}}(s) \\ T_{\text{l}}(s)
\end{bmatrix},
\end{equation}
where $\mathrm{H}(s)$ is defined as
\begin{equation}\label{eqn:H(s)}
\mathrm{H}(s)\!\! := \!\! \begin{bmatrix}
\!J_{\text{m}}s^2
\!+\!
(B_{\text{m}}\!+\!B_{\text{ml}})s
\!+\! 
K_{\text{s}}
& 
-B_{\text{ml}}s
-K_{\text{s}}
\\
- B_{\text{ml}}s - K_{\text{s}}
&\!\!\!\!\!\!\!
J_{\text{l}}s^2
\!+\!
(B_{\text{l}}\!+\!B_{\text{ml}})s
\!+\! 
K_{\text{s}}\!\!
\end{bmatrix}\!\!.\!\!
\end{equation}
The torque disturbance of the load is negligible due to designed structure. 
Accordingly, we obtain the following transfer functions from \eqref{eqn:Transfer_theta_tau} and \eqref{eqn:H(s)}, 
\begin{align}
T_1(s) &= 
\frac{\Omega_{\text{m}}(s)}{T_{\text{m}}(s)} = \frac{J_{\text{l}} s^2 + B_{\text{ml}} s + K_{\text{s}}}{\det\mathrm{H}(s)},\label{eqn:T1}
\\
T_2(s) &= \frac{\Omega_{\mathrm{l}}(s)}{\Omega_{\text{m}}(s)} = \frac{B_{\text{ml}} s + K_{\text{s}}}{J_{\text{l}} s^2 + B_{\text{ml}} s + K_{\text{s}}}, \label{eqn:T3}
\end{align}
where $\Omega_{\mathrm{m}}$ and $\Omega_{\mathrm{l}}$ are respectively the Laplace transform of $\omega_{\text{m}}$ and $\omega_{\text{l}}$.
For the transfer function between the voltage applied to the armature and the rotational velocity of the load \citep{2DOFControl1}, one can easily see
\begin{equation*} \label{eqn:G(s)_eq1}
\begin{split}
G(s) 
:= \frac{\Omega_{\text{l}}(s)}{V_{\text{a}}(s)} 
=K_{\text{t}}
\!
\big{(}K_{\text{t}}K_{\text{b}} \!+\!(L_{\text{a}} s \!+\! R_{\text{a}})T_1(s)^{-1}\big{)}^{-1}T_2(s),
\end{split}
\end{equation*}
from equations \eqref{eqn:omega_ml} and \eqref{eqn:T3}.
Since $K_{\text{s}} \gg 1 $, one can approximate $T_1(s)^{-1}$ by
$((J_{\text{m}}+J_{\text{l}}) s +B_{\text{m}})$ 
due to frequency range of the operation. %
Finally, we obtain
\begin{equation}
\label{eqn:11}
\begin{split}
G(s) &=  
\frac{K_{\text{t}}}{K_{\text{t}} K_{\text{b}} + \Big(L_{\text{a}} s+ R_{\text{a}}\Big)\Big((J_{\text{m}}+J_\mathrm{l}) s +B_{\text{m}}\Big)}
\\& \quad \quad 
\bigg({\frac{B_{m\mathrm{l}} s + K_{\text{s}}}{J_\mathrm{l} s^2 + B_{m\mathrm{l}} s + K_{\text{s}}}}\bigg).
\end{split}
\end{equation}

\subsection{The Control Scheme} \label{sec:control}
\begin{figure}[b]
	\centering
	\includegraphics[width=0.485\textwidth]{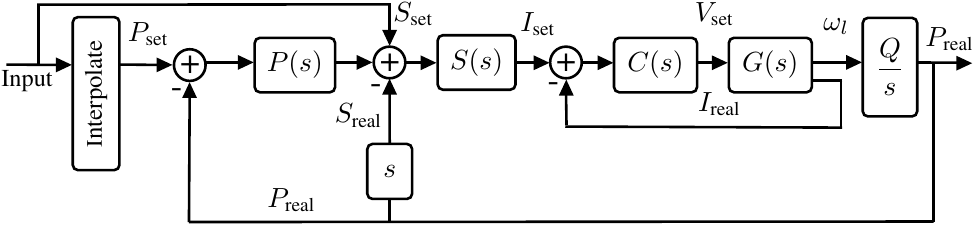}%
	\caption{Block diagram of the system}
	\label{fig:Block2}
\end{figure}
The system is controlled by a PLC that runs a custom-made software package named LASAL.
The controller consists of three cascaded loops as shown in Figure  \ref{fig:Block2}, where each loop regulates a different attribute of the system. In each control loop, the output signals serve as the reference for the next inner loop. 

The first block in the axis controller is the {\em interpolation} block which receives the trajectory specs from the user and determines the references for the position and the speed in the system. 
The interpolation block requires four inputs: the position set point, the speed set point, the desired acceleration and desired deceleration. 
Once these inputs are provided, the interpolation block generates a reference speed and position trajectory using the equations of motion.
The outer-most and middle control loops are respectively for the regulation of linear position and speed.
The output of the interpolation block provides these loops with the designed nominal references. 
The motor encoder detects the position of the motor and provide the feedback for both of these loops.
The controllers in the position control loop, denoted by $P(s)$, is a P-controller, whereas the controller in the speed control loop, denoted by $S(s)$, is a PI-controller. More precisely, we have $P(s) = K_\text{p}$ and $S(s) = K_\text{v} + \frac{K_\text{i}}{s}$. 
The speed control loop is followed by the current controller, which is the inner-most loop.
The feedback in this loop is the measured current of the armature.
This loop is regulated by a PID-controller block, denoted by $C(s)$, given as
$C(s) = K_\text{{cp}} + \frac{K_\text{{ci}}}{s}  + K_\text{{cd}} $. 
The output of the controller is the voltage set point for the motor which is regulated according to the set reference via a motor drive system converting the voltage reference to a corresponding input voltage. 
Finally, the last block 
is for conversion of the rotational velocity of the ball-screw to linear speed.

The linear axis has three separate modes of operations according to which the active control loops and parameters are chosen.
In position control mode (used in this work), all three feedback loops are active and the position is the most critical attribute of the system. 
In this mode, the controller will try to adhere as closely as possible to the position reference even if that entails deviating from the ideal speed trajectory.
Similarly, in the speed control mode, the speed trajectory is prioritized 
and the position controller deactivated by setting the gain in the position controller to zero, $K_\text{p}=0$. 
The third mode is the current control mode, in which only the innermost loop is active and the other controller gains are set to zero. 
\subsection{The Parameters of the Model}
The transfer function of the plant as well as the control loops depend on several parameters. 
Regarding the control loop, since we are only tuning the parameters of $P(s)$ and $S(s)$, it is assumed that the parameters of $C(s)$ are fixed and given.
Concerning the parameters of the plant, almost all of the values are provided in the available data sheets or can be calculated accordingly. The only exception here is $K_{\text{s}}$. 
We estimate this parameter by performing a simple experiment.
More precisely, we take the step response of the system first, and then, fit the step response of the model by fine-tuning parameter $K_{\text{s}}$ using least squares fitting. 
The resulting value as well as other known parameters are given in Table \ref{table:motor_Parameters}.

\begin{table}[t]
\caption{Weights for the cost function of the speed and the position controllers}
\centering
\ra{1.1}
\begin{tabular}[h]{ @{}c c c c  @{} }\toprule
\textbf{Parameter} &  \textbf{Value}\\\midrule
$K_\text{{cp}}$ & $60$ 				\\  
$K_\text{{ci}}$ & $1000$ 			\\  
$K_\text{{cd}}$ & $18$				\\   
$R_{\text{a}}$  & $9.02$ $\Omega$	\\  
$L_{\text{a}}$  & $0.0187$   		\\   
$K_{\text{t}}$  & $0.515$ Vs $\text{rad}^{{-1}}$ \\   
$K_{\text{b}}$  & $0.55$ Nm$\text{A}^{-1}$		 \\  
$J_{\text{m}}$  & $0.27\times 10^{-4} \text{ kg m}^2$  \\ 
$B_{\text{m}}$  & $0.0074$  	    \\  
$J_\text{l}$    & $6.53 \times 10^{-4}\text{kg m}{}^2$ \\  
$B_\text{{ml}}$ & $0.014$ 			\\  
$K_\text{s} $   & $3 \times 10^7$ 	\\
$Q$             & $1.8$ cm          \\ 
Maximum Speed   & $8000$ RPM  	    \\\midrule  
\end{tabular}
\label{table:motor_Parameters}
\end{table}   

\section{Numerical Experiments for Controller Tuning}\label{sec:numerical}

\subsection{Standard Tuning Methods} 
The classical PID tuning approach is the Ziegler-Nichols method, a heuristic designed for disturbance rejection \citep{ziegler1942optimum}.
PID auto-tuning technique is an automated version of Ziegler-Nichols method, the controller is replaced by a relay and the PID coefficients are estimated based the resulting oscillatory response of the system \citep{hang2002relay}.
Other tuning approaches are also used in practice, where a performance indicator of the system is optimized, for example the {\em integral of time-weighted absolute error} (ITAE) \citep{aastrom1993automatic}.
\subsection{Performance metrics and exaustive evaluation}

The main ingredient in Bayesian optimization \citep{BayesOpt3} is the cost function, which is composed of a set of metrics capturing the performance requirements of the system. For a linear actuator, the position tracking accuracy and the suppression of mechanical vibrations (oscillation effects) are of highest importance. 
A fundamental constraint on the controller gains is the stability. Here, it is achieved by constraining the ranges of the controller gains to known ranges: $ K_\text{v} \in (0,0.5], K_\text{i} \in (0,900], K_\text{p} \in (0,4200]$, derived from the numerical computation of the system response and the controller parameters. 

For the speed controller, the corresponding performance metrics extracted from the response of the system at different values of the controller gains are the overshoot, $\hs$, and the settling time of the speed step response, $\Ts$, as standard parameters for tuning. Furthermore, to quantify the performance of the system, the speed tracking error quantified by its infinity norm $\| \es\|_{\infty}$, and the integral of the time-weighted absolute value of the error of the speed response, $\es_{\text{ITAE}}$, are included, where $\text{s}$ indicates that the performance metric is associated with the speed controller. The latter is evaluated only after the motion is complete and the system speed set point is zero, and measures oscillations in the system due to excitation of vibrational modes.
The optimal gains 
are found following the minimization of the controller cost, which is given by the weighted sum of the performance metrics:
\begin{equation}\label{eq:cost_speed_sim}
\begin{array}{c}
\fs =  \sum_{k=1}^{\Ns} \gammas_k \Fs_{k}, \, 
\end{array}
\end{equation}
where $\Fs= [\Fs_k]_{k=1}^{\Ns} := [\Ts, \hs, \es_{\text{ITAE}}, \| \es\|_{\infty}]$, and $\Ns$ indicates the number of components in $\Fs $.

\begin{table}[t]
	\caption{Weights for the cost function of the speed and the position controllers}
	\centering
	\ra{1.3}
	\begin{tabular}[h]{ @{}l c c c c c @{} }\toprule
	\multicolumn{2}{c}{speed}  &&   \multicolumn{2}{c}{position}\\
	\cmidrule{1-3} \cmidrule{4-5}
	$\Fs_k$ & $\gammas_k$        && $\Fp_k$ & $\gammap_k$       \\ \midrule 
	$\Ts$ & $500$                && $\Tp$ & $10^4$              \\ 
	$\hs$ & $2$                  && $\hp$ & $10$                \\ 
	$\es_{\text{ITAE}}$ & $10^4$ && $\hps$ & $15$               \\  
	$\|\es\|_{\infty}$ & $500$   && $\|\es\|_{\infty}$ & $100$  \\ \bottomrule
	\end{tabular}
	\label{table:weights_grid}
\end{table}   

\begin{figure}[b]
	\centering
	\includegraphics[width=0.49\textwidth]{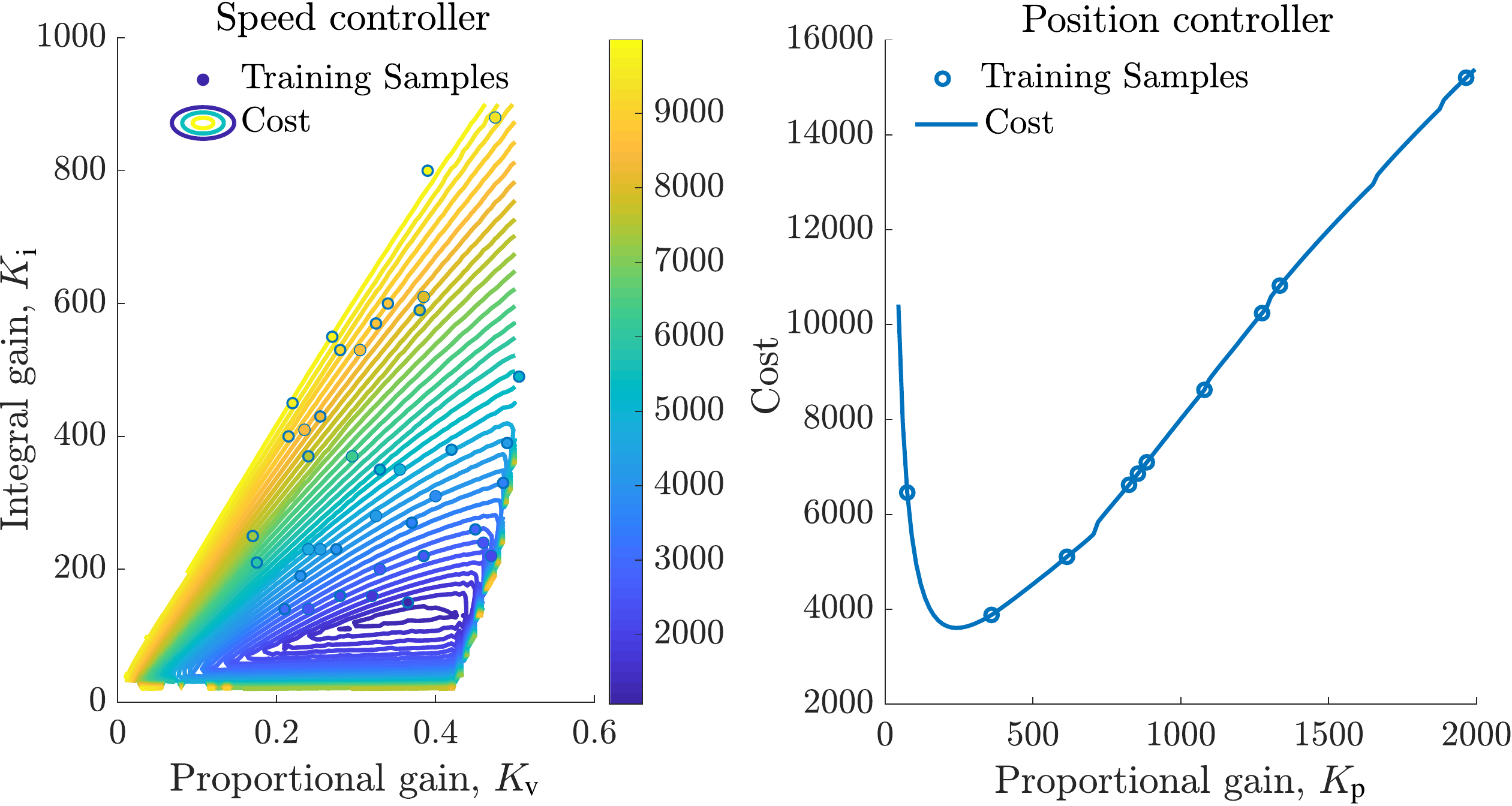}
	\caption{Cost function and training points corresponding to one sampling for the speed controller (left) and the position controller (right)}
	\label{fig:costs}
\end{figure}

\begin{figure*}[ht]
	\centering
	\includegraphics[width=0.725\textwidth]{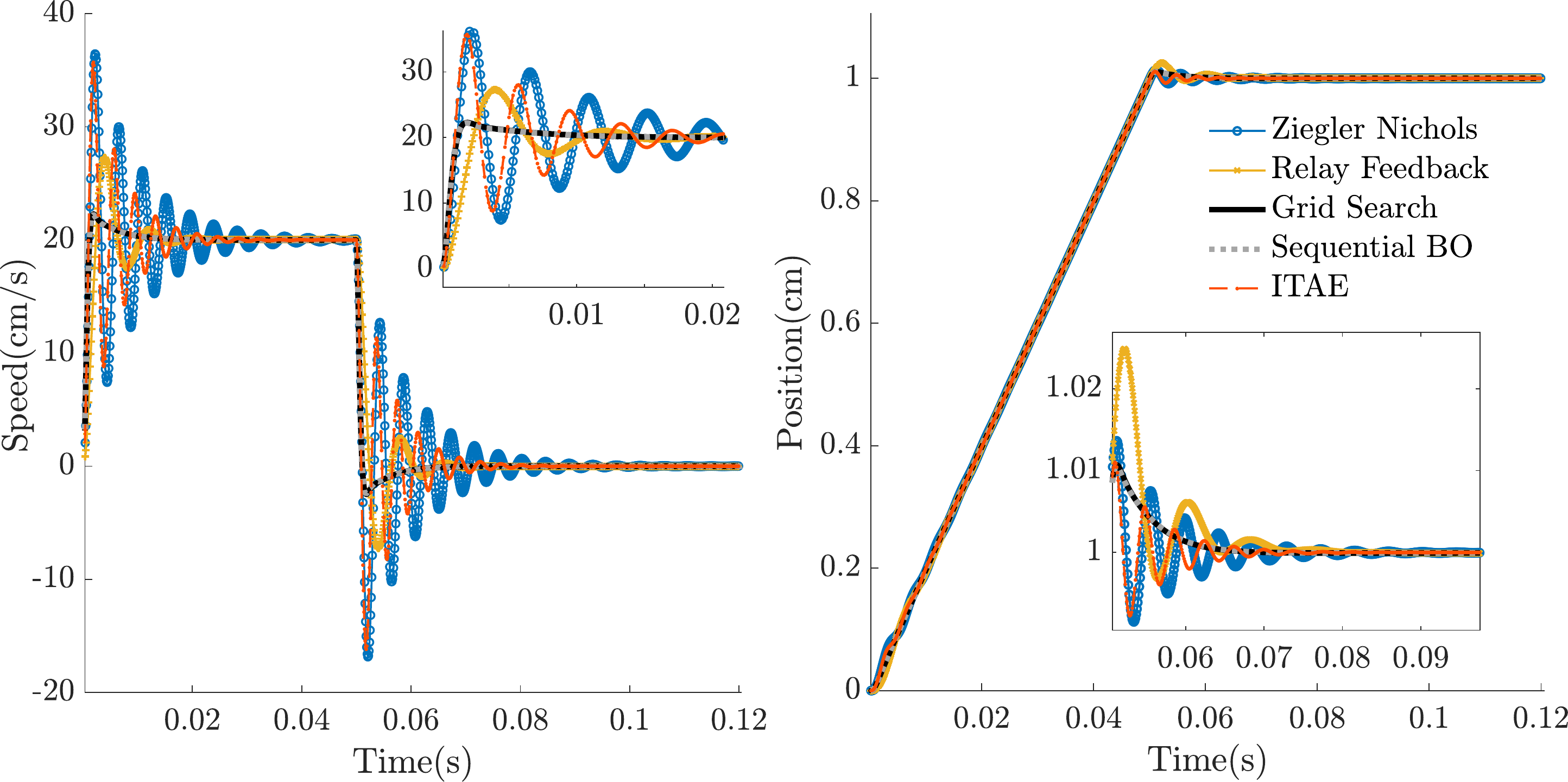}
	\caption{Speed and position response for different benchmark tuning methods}
	\label{fig:benchmark}
\end{figure*}
The gains corresponding to the minimal cost following evaluation of the cost function for all combinations of gains on a grid are found to be $K_\text{v} = 0.36$, $K_\text{i} = 130$, as shown in Table \ref{table:SimCompare}. For the explored ranges of the controller gains the grid spacing is $10$ for $K_\text{i}$, $0.005$ for $K_\text{v},$ and $15$ for $K_\text{p}$, for grid search and for Bayesian optimization.
Figure \ref{fig:costs} shows that the optimal region where the cost for the controller gains in speed control mode is minimal is rather flat and the same performance can be achieved for $K_\text{v} \in [0.36,0.39]$ and $K_\text{i} \in [90,130]$.

\begin{table}
	\caption{Controller gains resulting from different tuning methods}
	\label{table:SimCompare}
	\centering
	\ra{1.3}
	\begin{tabular}[h]{@{}l c c c c  @{}}\toprule	
		Tuning method  &  $K_\text{p}$ &  $K_\text{v}$ &  $K_\text{i}$\\ \midrule
		Grid search (optimal value)&$225$&$0.36$&$130$\\ 
		Ziegler Nichols &$392$&$0.18$&$510$ \\  
		ITAE criterion&$255$&$0.11$&$420$ \\ 
		Relay tuning&$115$&$0.05$&$130$ \\ 
		Sequential Bayesian optimization &$225$&$0.37$&$130$\\ \bottomrule  
	\end{tabular}
\end{table}   

The optimal $K_\text{p}$ is found by setting the speed controller parameters to the optimized values ($K_\text{v} = 0.36, K_\text{i} = 130$), which are found by the BO tuning of the speed controller, and coincide as well with the gains determined by grid search, as shown in Table \ref{table:SimCompare}. The corresponding position controller cost function is then evaluated for varying values of $K_\text{p}$ for this specific speed controller.
For the system in position control mode, the corresponding performance metrics extracted from the response of the system are the overshoot $\hp$ and the settling time $\Tp$ of the position step response, the tracking error in position quantified by its infinity norm $\| \ep\|_{\infty}$, and the overshoot of the actual speed $\hps$, where $\text{p}$ indicates that the performance metric is associated with the position controller. 
The latter is a measure of the effect of the position controller gain on the speed of the system.
The cost function used to find the optimal position controller according to the performance metrics is 
\begin{equation}\label{eq:cost_seq_post}
\begin{array}{c}
\fp = \sum_{k=1}^{\Np} \gammap_k \Fp_{k}, \, 
\end{array}
\end{equation}
where $\Fp= [\Fp_k]_{k=1}^{\Np} := [\Tp, \hp, \hps, \| \ep\|_{\infty}]$. 

Note that in both of \eqref{eq:cost_speed_sim} and \eqref{eq:cost_seq_post}, the weights are chosen due the order of magnitude and importance of correspond performance metric. 

The cost functions for the speed and the position controllers are shown in Figure \ref{fig:costs}, with a grid spacing $10$ for $K_\text{i}$, $0.005$ for $K_\text{v},$ and $15$ for $K_\text{p}$. 
The optimal position controller found by grid evaluation is $K_\text{p} = 225$, where the grid parameters are same as above.

\subsection{Sequential Bayesian Optimization}
After defining the corresponding cost functions, they can be modelled using GP regression and used in Bayesian optimization to find optimal controller gains. To increase the accuracy of the models, we first collect data at random locations of the controller gains to form a prior distribution of the GP models. The number of training samples in this phase has a direct influence on the number of iterations needed to reach a stopping criterion that defines the converged controller gains.

Here, initially the speed controller gains $K_\text{v}$ and $K_\text{i}$ are tuned, without connecting the position controller. Once the optimal speed controller gains are found, $K_\text{p}$ is tuned while keeping the speed controller fixed at the optimal gains. 
Following a selection of a number of random configurations of $K_\text{v}$ and $K_\text{i}$, the prior cost, calculated using \ref{eq:cost_speed_sim} is modelled with a GP, and the acquisition function is minimized by grid search to predict the next plausible configuration of $K_\text{v}$ and $K_\text{i}$ where the cost should reach a lower value. At this candidate configuration, the model of the cost function is updated, and the procedure is repeated.
Following several iterations (summarized in Table \ref{table:trainIterSeq}), the optimization terminates within a narrow set of optimal values confined within the flat region of the cost minimum, as shown in Figure \ref{fig:costs}. Depending on the initial GP model of the cost function, and on the initial number of measurements, the number of iterations needed to reach convergence changes.

Once the optimal values of $K_\text{v}$ and $K_\text{i}$ are found, they are kept fixed and the position controller gain $K_\text{p}$ is in turn optimized by minimization of the position controller cost modeled by Gaussian process regression. The algorithm is initialized by randomly selecting a number of inputs for training data, and the maximum number of iterations is set to be 20. The optimization algorithm terminates in 3-6 iterations (depending on the number of  training data used for the prior, see Table \ref{table:trainIterSeq}), and the resulting position controller gain is $K_\text{p} = 225$. The corresponding system response is shown on Figure \ref{fig:benchmark}, with the position controller with a set speed of 100 $\text{cm/s}$ for the position controller, and a set position of 60 $\text{cm}$ for the speed controller. Both the response traces and the tuned controller gains closely match those corresponding to the grid simulations, as shown in Table \ref{table:SimCompare} and on Figure \ref{fig:benchmark}. In the speed control mode, the speed performance corresponding to BO tuning shows an extremely low overshoot, and settles quickly to the nominal value. In the position mode, the position tracking delay and overshoot are significantly reduced which is crucial as a high overshoot in position can cause the machine to hit the edges and activate the limit switches, which switches off the motor and results in an error state. The position error in steady state mode is minimized and the effect of position gain on the speed response is taken into consideration, thereby resulting in a small increase in the speed response overshoot. This response is significantly improved with respect to standard tuning methods as shown in Figure \ref{fig:benchmark}, and has the lowest overshoot, settling time, and position or speed errors. 
Table \ref{table:SimCompare} shows that the result of Bayesian optimization tuning is closest to the exhaustive evaluation results obtained on a grid. The values of the gains obtained via standard tuning approaches (Zeigler-Nichols, relay tuning, and ITAE) are more aggressive and show significantly higher overshoot and oscillations, as shown in Figure \ref{fig:benchmark}.
\begin{figure*}[t]
	\centering
	\includegraphics[width=0.725\textwidth]{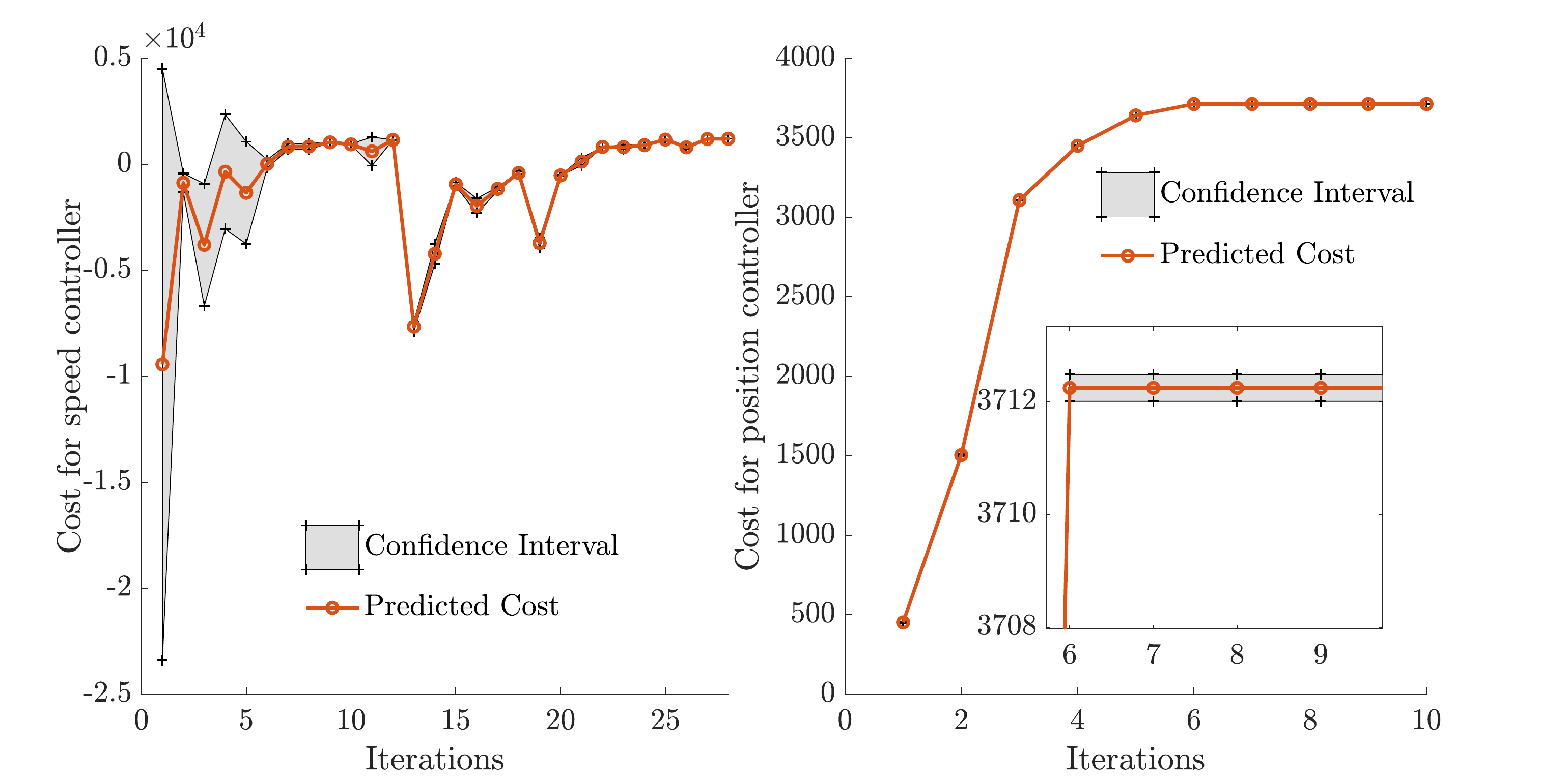}
	\caption{Predicted cost and associated uncertainty for the sequential BO. The inset in the right panel shows a close up on the confidence intervals in the predicted mean of the cost for the position controller gain.}
	\label{fig:BOsim}
\end{figure*}

\begin{table}[b]
	\caption{Effect of training data size $N_{\text{train}}$, on the number of required iterations of sequential BO, $N_{BO}.$}
	\label{table:trainIterSeq}
	\centering
	\ra{1.3}
	\begin{tabular}[h]{@{}l c c c c c c c c @{}} \toprule
		\multicolumn{4}{c}{speed} & &  \multicolumn{3}{c}{position}\\
		\cmidrule{1-4} \cmidrule{6-8}
		$N_{\text{train}}$ &  $N_{\text{BO}}$  &  $K_\text{v}$ &  $K_\text{i}$ && $N_{\text{train}}$ &  $N_{\text{BO}}$  &  $K_\text{p}$\\ \midrule
		50&19&0.37&130 && 15&3&225  \\ 
		30&27&0.345&130 && 10&6&240 \\ 
		20&44&0.36&110 && 7&5&210 \\ \bottomrule
	\end{tabular}
\end{table}   

The optimal parameters of the controller can be found following an initial exploration phase that requires data collection at 30-40 different configurations of parameters, and a tuning phase that requires 20-30 iterations in total. As the performance metrics can be fully automated, and the initial exploration phase needs to be repeated only upon major changes in the system, the proposed tuning method can be efficiently implemented. The evolution of the cost prediction for each subsequent iteration and the associated uncertainty are shown in Figure \ref{fig:BOsim}, for 30 training points. Initially the uncertainty is very high and the predicted mean of the cost is negative. According to the termination criterion, the optimization terminates after a minimum in the cost is repeated more than three times. Accordingly, a drop in the variance is observed around these values. The low uncertainty of the cost of the position controller gain can be explained by the sufficiently high number of points used to calculate the prior and indicates that the training data can be reduced. 
The proposed BO tuning thus offers a trade-off between grid based search, and heuristic-based methods. Grid search requires extensive number of experiments to evaluate all parameter combinations, and provides the optimal gains (according to a set criterion), whereas standard tuning methods require significantly reduced number of experiments, but often result in conservative gains. With BO tuning a relatively small number of experiments leads to the optimal gains, specified according to the data-driven optimization objective and termination criterion.

\section{Conclusion and Outlook} \label{sec:conclusion}
In this paper, a data-driven approach for cascade controller tuning based on Bayesian optimization has been demonstrated in simulation. It enables fast and standardized tuning, with a performance superior to other auto-tuning approaches. Furthermore, it enables easy adaptation of the controller parameters upon changes in the load, or in the mechanical configuration of the system. Extending the method with automatic detection of instabilities \citep{ConstrainedBOCascade}, or safe exploration in evaluation the cost will further extend its flexibility and potential for practical use.
\section{Acknowledgment}
The authors gratefully acknowledge Piotr Myszkorowski and Sigmatec AG who provided technical assistance with the LASAL software, as well as the linear axis drive system and the associated PLC.
\bibliography{bibliography}
\iftrue 
	\appendix
\section{Gaussian Processes and Bayesian Optimization}
Bayesian optimization (BO) is a data-driven approach for solving optimization problems with unknown objective function.
More precisely, the objective function is available only in form of an {\em oracle}. 
BO efficiently samples and learns the function on-line by querying the oracle, and subsequently, finds the global optimum iteratively. It uses Gaussian process regression (GPR)  to build a surrogate for the objective and quantify the associated uncertainty~\citep{BayesOpt4}. In each iteration, the data is used for deciding on the next evaluation point based on a pre-determined acquisition function. The new information gathered is combined with the prior knowledge using Gaussian process regression to estimate the function and find the minimum~\citep{BayesOpt4}.
One of main advantages of BO is its potential in explicitly modeling noise which is automatically considered in the uncertainty evaluation without skewing the result~\citep{BayesOpt1}.

\subsection{Gaussian Processes} \label{GP}
A Gaussian process (GP) is a collection of random variables where each of its finite subset is jointly Gaussian \citep{GPR}. 
The GP is uniquely characterized by the mean function, $\mu:\Xcal$, and covariance/kernel function, $k:\Xcal\times\Xcal\to \Rbb$, where $\Xcal$ is the set of location indices.
Accordingly, the GP is denoted by $\GP{\mu}{k}$.
Commonly in literature, $\Xcal=\Rbb^d$ and $k$ is {\em square exponential} kernel which is defined as
\begin{equation}
k_{\text{SE}}(x,x') = \sigma_{f}^2\exp{(-\frac{1}{2}(x\!-\! x')^{\tr}\mathrm{L}^{-1}(x\!-\!x'))},\  \forall x,x'\!\in\Rbb^d,
\end{equation}
where $\sigma_f$ and $\mathrm{L}$ are the hyperparameters of the kernel respectively referred as {\em flatness parameter} and {\em length scale matrix}.

Gaussian processes provide suitable flexible classes for Bayesian learning by introducing prior distributions  over the space of functions defined on $\Xcal$.
In fact, due to the favorable properties of the Gaussian distributions, the marginal and conditional means and variances can be computed on any finite set of locations in a closed form. 
Subsequently, a {\em probabilistic  non-parametric regression}  method can be developed \citep{GPR}.
More precisely, let $f:\Xcal\to\Rbb$ be an unknown function, with $\GP{\mu}{k}$ prior and noisy measurements at $N$ training location indices $\mathrm{x} = [x_1, x_2, ...x_N]$ as  $\mathrm{y} = [y_1, y_2, ...y_N]$, i.e., $y_i=f(x_i)+n_i$, where $n_i$ is the measurement noise with distribution $\Ncal(0,\sigma_n^2)$, for $i=1,\ldots,N$ and $f$. 
For a new location $x$ where the corresponding measurement is not provided, since the joint distribution of the measurements data is a Gaussian with a given mean and a given covariance, one can predict the value of  measurement at the new location. To this end, define the Gram matrix, $\mathrm{K}_{\mathrm{xx}}$, where its  element at the $i^{\text{\tiny{th}}}$ row and the $j^{\text{\tiny{th}}}$ column is given by $k(x_i,x_j)$. Then, we have
\begin{equation}
     \mathrm{y} \sim \mathcal{N}\bigg{(}0,\mathrm{K}_{\mathrm{xx}}+\sigma^2_n\mathbb{I}\bigg{)} \,.
\end{equation}  
The hyperparameters can be estimated by minimizing the negative marginal log-likelihood of the joint distribution of the training data, i.e., given that $\theta\in\Theta$ is the vector of hyperparameters, one can estimate $\theta$ by
\begin{equation}\label{eqn:min_nlml}
\hat{\theta}:= 
\argminOp_{\theta\in\Theta}\ -\log p(\mathrm{y}|\mathrm{x},\theta) \, ,
\end{equation}
where $ p(\mathrm{y}|\mathrm{x},\theta)$ is the probability density function of the labels or measurements acquired at locations $x$. 
The joint distribution of the training data with a new data point, with an unknown label $y_x = f(x)$, can be calculated as follows:
\begin{equation}
    \begin{bmatrix}
    \vc{y}\\y_x
    \end{bmatrix} \sim \mathcal{N}\bigg{(}0,\begin{bmatrix} \mathrm{K}_{\mathrm{xx}}+\sigma^2_n\mathbb{I} & \mathrm{k}_{\mathrm{x}x}^\tr \\ \mathrm{k}_{\mathrm{x}x} & k_{xx} \end{bmatrix}\bigg{)}  \,.
\end{equation} 
where $\mathrm{k}_{\mathrm{x}x}\in\Rbb^N$ is a vector which its $i^{\text{\tiny{th}}}$ element is given by the kernel as $k(x,x_i)$, for any $i=1,\ldots,N$, and $ k_{xx}:=k(x,x)$. Accordingly, the posterior distribution of $y_x|\mathrm{y}$ is  a Gaussian as
\begin{equation}
    y_x | \mathrm{y} \sim \mathcal{N}(\mu(x),\sigma(x)) \,,
\end{equation}
where the mean of prediction, $\mu(x)$, and the corresponding covariance, $\sigma(x)$, are given as 
\begin{align}
\mu(x) &:= \mathrm{k}_{\mathrm{x}x}^\tr\big{(}\mathrm{K}_{\mathrm{xx}}+\sigma^2_n\mathbb{I}\big{)}^{-1}\mathrm{y},
\label{eqn:mu}\\
\sigma(x) &:= k_{xx} - \mathrm{k}_{\mathrm{x}x}^\tr\big{(}\mathrm{K}_{\mathrm{xx}}+\sigma^2_n\mathbb{I}\big{)}^{-1}\mathrm{k}_{\mathrm{x}x}. \label{eqn:sigma}
\end{align}
One can see that $\mu(x)$ is a nonlinear function predicting the value of $f$ at location $x$ with an uncertainty  described by $\sigma(x)$. Accordingly, this is a nonlinear regression method called {\em Gaussian process regression} (GPR).
\subsection{Bayesian Optimization Algorithm}
Toward finding the optimum parameters of the controller, BO uses GPR iteratively to reduce the uncertainty and improve the result at each step. It is a technique based on randomness and probabilistic distribution of an underlying objective function that maps the optimization parameters to a user-defined cost function. The optimization starts with some initial data and a prior distribution mean and covariance which captures the available knowledge about the behaviour of the function. We use GPR to update the prior and form the posterior distribution mean and variance over the objective function. The posterior distribution is then used to evaluate an acquisition function that determines the location of the subsequent candidate point, denoted by $x_n$, at iteration $n$. Data collected at this new candidate location is appended to the previous data to form a new training set for the GPR of the defined objective. The process is repeated with the new training data set and the posterior is updated with each added data point. This cycle continues until a termination criterion is fulfilled. 

The acquisition function is a sampling criterion that determines the sampling of subsequent candidate points and varies depending on system requirements. Instead of directly optimizing the expensive objective function, the optimization is performed on an inexpensive auxiliary function, the acquisition function, which uses the available information from the GP model in order to recommend the next candidate point $x_{n+1}$. Commonly used acquisition functions include \textit{entropy search}, \textit{expected improvement}, \textit{upper confidence bound (UCB)}, etc \citep{BayesOpt4}. The acquisition function provides a trade-off between exploitation and exploration. It can explore regions of the domain with highest prediction uncertainty, exploit the point where the cost is predicted to be lowest, or select new locations according to a combination of the two objectives\citep{BayesOpt3}. In the controller tuning problem explored here, the goal is to minimize the cost function defined through performance indicators based on the data obtained at each candidate configuration of parameters, and to find the parameters that achieve this minimum. The acquisition function used for this system is based on the UCB acquisition function  which uses the upper bound to maximize the cost function, whereas we use the lower bound in order to minimize the cost (\textit{Lower confidence bound, LCB}). 

The LCB acquisition function from  \citep{BayesOpt3} is 
\begin{equation}
\label{eqn:6.6}
x_{n+1}=\argminOp_{x \in \Dcal} \mu_{n}(x)-\beta_{n}\sigma_{n}(x) \, ,
\end{equation}

$\beta_{n}$ is a constant that specifies the confidence interval around the mean should be considered, and $\Dcal$ is the allowed range of the optimization variables, in this case the range of gains where the controllers are stable. This objective prefers both points $x$ where $f$ is uncertain (large $\sigma_{n}$) and such where we expect to achieve lowest cost ($\mu_{n}(\cdot)$). It implicitly negotiates the exploration - exploitation trade-off. A natural interpretation of this sampling rule is that it greedily selects points $x$ where $f(x)$ could be lower than the current minimum and upper bounds the function by the current minimum \citep{BayesOpt3}.

The termination criteria for the BO algorithm need to be decided based on application and system specifications. In this application, one of the criteria used is that the number of iterations should be limited to $N_{max}$ since the primary goal is to reduce the number of evaluations on the system. In addition, another termination criterion used is repeated sampling around the current minimum.

\fi 
\end{document}